\documentclass[twocolumn,showpacs,preprintnumbers,amsmath,amssymb]{revtex4}

\usepackage{graphicx}
\usepackage{dcolumn}
\usepackage{bm}

\begin{document}

\preprint{APS/123-QED}

\title{Excitation spectrum of hydrogen adsorbed to carbon nanotubes}

\author{B. Renker,$^{1}$ H. Schober,$^{2}$ P. Schweiss,$^{1}$
       S. Lebedkin,$^{3}$ and F. Hennrich$^{4}$}
\affiliation{$^{1}$Forschungszentrum Karlsruhe, IFP, P.O.B. 3640,
       D-76021 Karlsruhe, Germany}
\affiliation{$^{2}$Institut Laue-Langevin, BP 156 X, F-38042
       Grenoble Cedex, France }
\affiliation{$^{3}$Forschungszentrum Karlsruhe, INT, P.O.B. 3640,
       D-76021 Karlsruhe, Germany}
\affiliation{$^{4}$Institut f\"ur Physikalische Chemie II,
Universit\"at Karlsruhe,
  D-76128 Karlsruhe, Germany}

\date{\today}

\begin{abstract}
We have studied the microscopic dynamics of hydrogen adsorbed to
bundles of single walled carbon nanotubes using inelastic neutron
scattering. Evidence is obtained for much higher storage
capacities in chemically treated compared to as prepared
material. This indicates an additional adsorption layer inside
the tubes. Well pronounced excitations in the H$_2$ spectrum at
low energies confirm this conclusion. The desorption of hydrogen
is monitored in real time as a function of temperature. Hydrogen
storage is highly stable below 150\,K in agreement with the
harmonic evolution of the hydrogen spectrum, which indicates a
strong binding potential. Above 200\,K hydrogen can be  released
in a controlled way by simple heating. The excitation spectrum
changes significantly during the release. Remnants of hydrogen
persist up to 400\,K.

\end{abstract}

\maketitle

Public interest in high performance storage devices has triggered
many detailed investigations of hydrogen adsorbed to graphite
surfaces. Besides a commensurate phase several incommensurate
phases were found at higher pressures \cite{Dillon,Nielsen}. H$_2$
storage capacities of activated carbon, however, seem to fall
considerably short of 6.5\,wt\%, which is the declared target
value of the United States Department of Energy \cite{Dillon}.
There is hope that single walled carbon nanotubes (SWNT's) may
approach this limit. Measurements via volumetric methods revealed
in the majority of cases disappointing storage capacities
$<$\,2wt\%, although confusing results reaching up to 17\,wt\%
exist in the literature. Recent reviews of the subject can be
found elsewhere \cite{Dillon,Hirscher}.

Fundamental insight into the physics of the adsorption process is
needed to identify ideal storage media. Understanding the
dynamics is as important as determining the structure. Due to the
exceptionally high incoherent neutron scattering cross-section of
hydrogen ($\sigma$({\rm H})\,=\,81\,barn) inelastic neutron
scattering (INS) is the method of choice for studying its
microscopic dynamics. Surprisingly, corresponding results are
rarely found in recent literature
\cite{Renker,Brown-CPL-2000,Narehood}. In this paper we will
present temperature dependent susceptibilities and generalized
phonon densities of states $G(\omega)$ for H$_2$ adsorbed by
bundles of SWNT's in the technologically interesting range above
77\,K. We thus do not cover contributions from hydrogen molecules
adsorbed on the surface of SWNT-bundles which desorb well below
77\,K as identified in a study of quantum rotations
\cite{Brown-CPL-2000,Narehood}. Measurements were performed on as
prepared (closed tubes) and chemically treated materials (opened
tubes). Some of our Raman (R) scattering results will be
discussed in this context.

Our SWNT's were produced by pulsed laser vaporization of carbon
rods doped with Co and Ni catalysts in a stream of Ar. The raw
material (bundles of SWNT's) contained up to 50\% SWNT's. For
purification the material was soaked in 3M HNO$_3$ and had after
filtration the consistency of ``bucky" paper. In all steps we
have essentially followed the procedures described in the
literature \cite{Rinzler,Bandow,Lebedkin}. Our SWNT samples of
typically 0.5\,g were in situ loaded with 1 bar of H$_2$ after
annealing for $\approx$\,24\,h at 750\,K in dynamic vacuum. Then
the samples were quickly transferred into a pressure cell and
loaded with H$_2$ at 200\,bar. Subsequently the samples were
quickly removed, filled into Al-containers and stored at 77\,K.
We have prepared several batches of samples all giving very
similar spectroscopic results. INS measurements were performed on
the IN6 time-of-flight spectrometer at the high flux reactor of
the Institut Laue Langevin in Grenoble, France. The Al-cylinders
with the samples were mounted in a cryofurnace suitable to adjust
temperatures between 77\,K and 400\,K. A small hole in the
containers allowed an escape of H$_2$ into dynamic vacuum.
Detailed Measurements were performed with an incident neutron
energy of 4.75\,meV in the upscattering mode at 150\,K, 300\,K,
400\,K and again at 300\,K and 150\,K.  The total counting time
was subdivided into steps of 0.5\,h in order to monitor the
desorption of H$_2$ as a function of temperature. Multiphonon
corrections were applied in a self-consistent way when
calculating the generalized density-of-states $G(\omega)$.

\begin{figure}
\includegraphics[width=0.8\linewidth]{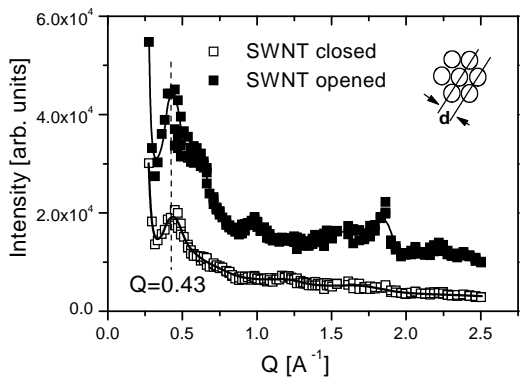}
\caption{Neutron diffraction spectra of as prepared (open
         squares) and chemically treated (full squares) SWNT's loaded with
         H$_{2}$. The first maximum corresponds to the (1 0) reflection
         from planes in a triangular lattice with a spacing of
d\,=\,14.6\,\AA.}
         \label{fig1.eps}
\end{figure}

\begin{figure}
\includegraphics[width=0.8\linewidth]{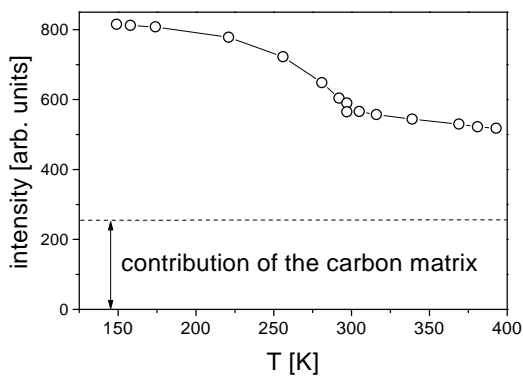}
\caption{Hydrogen desorption for the chemically treated material
         monitored by the integrated elastic scattering intensity.
         Below 300\,K weakly bond H$_2$ inside the tubes desorbs.
         A much slower release is observed at higher temperatures.
         Chemically active sites may provide docking
         centers important for this range. A contribution of the
         carbon matrix is estimated from the intensity observed
         for the as prepared material at the end of the cooling cycle.}
         \label{fig2.eps}
\end{figure}

In Fig.~1 we show diffraction data of our hydrogen loaded samples
normalized to the same mass as obtained from the angular
dependence of the elastic scattering intensity at 150\,K on IN6.
The first maximum at $ Q = 0.43$\,\AA$^{-1}$ is attributed to
reflections from $(1 0)$ planes of triangularly arranged tubes
with a lattice spacing of 14.6\,\AA . Using a commonly accepted
intertube spacing  of 3.2\,\AA\ \cite{Rols2} this leads to
individual tube diameters of $d\approx 13.67$\,\AA . This value
is close to that of $(10,10)$ armchair tubes and in excellent
agreement with both our Raman results\cite{foot1} and data
published in literature for the technique of laser ablation and
the presently used catalysts \cite{Rinzler,Bandow}. From TEM
images of our sample material we conclude on numbers
of\;$\approx$\,30 SWNT's in a bundle. The width of the main $(1
0)$ diffraction peak is mainly caused by the finite diameter of
the bundles. Because of the crude Q-resolution of the IN6
spectrometer which is optimized for inelastic measurements we
will not perform any  more detailed analysis. No principal
changes occur for the chemically treated material although there
are indications for some modifications in the amorphous part of
the material (the second small maximum at $Q \approx
1.9$\,\AA$^{-1}$ \cite{Rols1}). The higher background level is
attributed to a considerably larger incoherent scattering
contribution of adsorbed hydrogen in the open tubes.

The evolution of hydrogen desorption with temperature is shown in
Fig.\,2 for the chemically treated sample. The intensity observed
for the as prepared material at the end of the measuring cycle is
taken as contribution of the carbon matrix. Starting at about
200\,K we observe for both types of samples a significant
decrease in the integrated elastic intensity. The step at 300\,K
is due to 3\,h's of measurement at this temperature. In the case
of as prepared material the intensity does not change upon
further heating, while a continuous evolution up to our maximum
temperature of 400\,K is observed for the chemically treated
sample. A significant rest of H$_2$ is still present at 400 \,K.
By comparison to a Vanadium standard we find a typical release of
$\approx$\,0.6\,wt.\% for the chemically treated samples. Almost
negligible are corresponding values for the as prepared material,
i.e. $\approx$\,0.05\,wt.\%. Based on the estimated  contribution
of the carbon matrix we conclude on a total H$_2$ adsorption
capacity for our samples with the opened tubes on the order of
1.2\,wt.\%.

\begin{figure}[t!]
\includegraphics[width=0.8\linewidth]{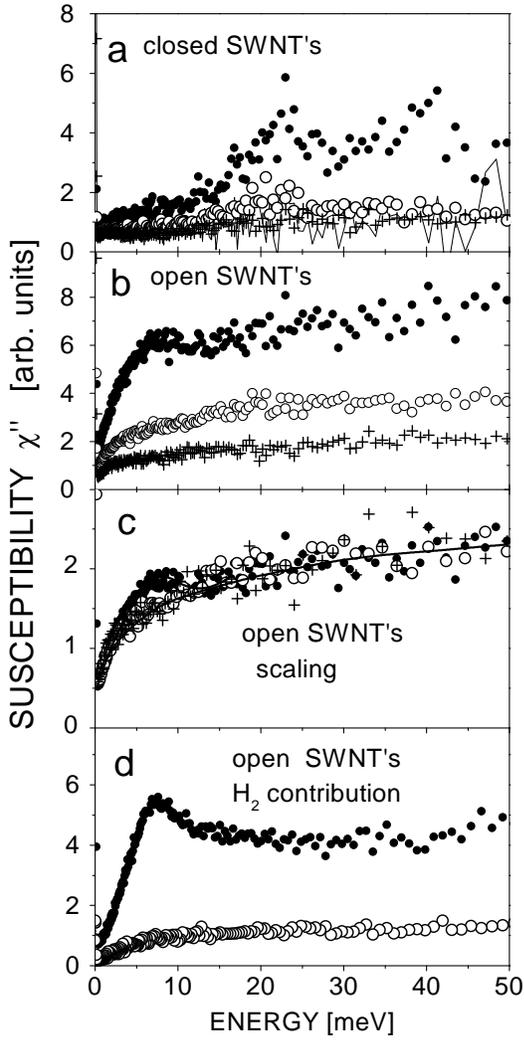}
\caption{The imaginairy part of the dynamical
         susceptibility $\chi\prime\prime$($\omega$)) measured at
         150\,K (full circles), 300\,K
         (open circles) and 400\,K (crosses) for a: closed SWNT's. The
         thin line is the result after the heat treatment with no hydrogen
         left. b: opened SWNT's. The adsorption of additional hydrogen is
         registered. c: opened SWNT's. It is possible to account for the
         desorbed hydrogen by application of a scaling factor
         (150\,K: 0.23; 300\,K: 0.56). d: opened SWNT's. The hydrogen
         contribution calculated from the difference of spectra.
         Results for 150\,K (full dots) and 300\,K (open dots) are shown.}
         \label{fig3.eps}
\end{figure}

Fig.\,3 shows the imaginary part of the dynamical susceptibility
$\chi^{\prime\prime}(\omega)$ as a function of temperature for
both kinds of sample. The spectra have been integrated over the
scattering angle to achieve better statistics.
$\chi^{\prime\prime}(\omega)$ is obtained from the measured
spectra by correcting for the thermal occupation factor. For a
regular, harmonic solid $\chi^{\prime\prime}(\omega)$ should be
temperature independent as Debye-Waller and multi-phonon terms
cancel each other to a large extent. This is actually what we
observe at low temperatures implying that the hydrogen is solidly
attached to the substrate. There are in particular no indications
of quasi-elastic scattering as expected for H$_{2}$ molecules in
the gas state. Apart from a remarkable peak around 7.5\,meV, which
is observed only at $T<300$\,K for the open SWNT's, the loss of
intensity in $\chi^{\prime\prime}(\omega)$ is found to be
proportional to the spectral distribution itself for nearly all
frequencies, i.e.\ consecutive $\chi^{\prime\prime}(\omega)$ can
be scaled onto each other via a common factor (Fig.\,3c). This
line-shape invariance indicates that we are not dealing with a
change in the dynamics, which would lead to a redistribution of
spectral weight. At 150\,K  $\chi^{\prime\prime}(\omega)$ for the
as prepared material (Fig.\,3a) shows strong similarities to the
excitation spectrum of the SWNT Matrix. This can be understood by
assuming that within this frequency region the H$_2$ molecules
vibrate in phase with the carbon atoms of the wall to which they
are attached. In the chemically treated material these well
structured contributions are completely dominated by much larger
contributions from H$_2$ adsorbed inside the tubes (Fig.\,3b). By
taking the difference of spectra for equivalent temperatures of
the heating and cooling cycle we are able to extract the
contribution of hydrogen desorbed in between the two measurements
(Fig.\,3d). The particular excitation at 7.5\,meV found in the
chemically treated material shows up clearly at 150\,K while no
such peak is found at 300\,K.

\begin{figure}
\includegraphics[width=0.9\linewidth]{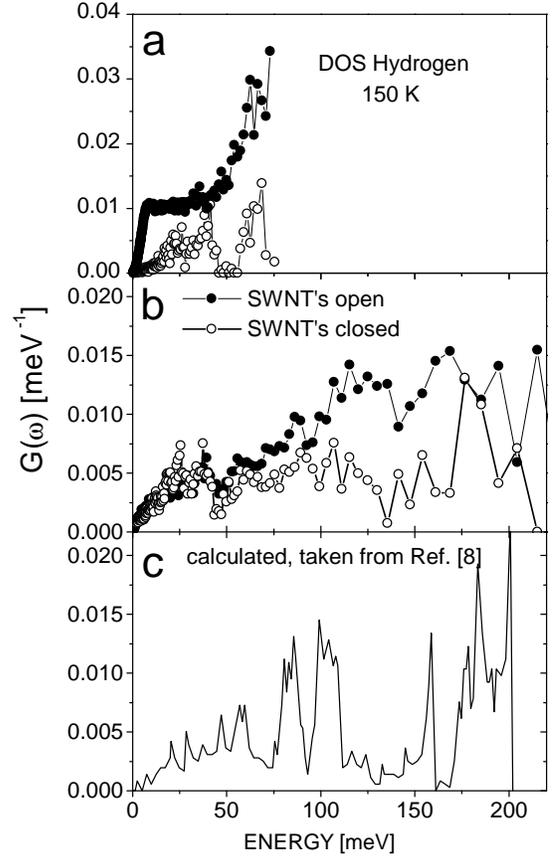}
\caption{The phonon density of states $G(\omega)$ (see text).
         Open and full circles refer to as prepared and chemically
         treated samples. a: the hydrogen contributions present
         at 150\,K. b: results obtained at 400\,K. Since $G(\omega)$
         for the closed tubes corresponds well to that one of the
         SWNT matrix shown in Fig.\,4c it is concluded that
         essentially all of the hydrogen is desorbed.
         The higher $G(\omega)$ observed for the open tubes is
         attributed to hydrogen chemisorbed by dangling bonds.}
         \label{fig4.eps}
\end{figure}

To gain further insight into the hydrogen dynamics we determined
the generalized densities-of-states $G(\omega)$ (Fig.\,4)\
\cite{foot2}. For 400\,K (Fig.4b) the frequency distribution of
the closed tubes compares very favorably to that measured and
calculated for a SWNT matrix \cite{Rols1} shown in Fig.\,4c. In
particular a grouping into high frequency tangential and low
frequency radial modes can be observed. The very low energy
excitations are due to inter-tube modes. The prominent radial
breathing mode shows up as a smaller peak at 28\,meV. The
hydrogen remaining in the chemically treated material gives rise
to considerable spectroscopic contributions still at 400\,K.
Fig.\,4a shows $G(\omega)$ of the isolated H$_2$ contribution for
as prepared and opened SWNT's at 150\,K in the experimentally
accessible energy region. The spectral distribution of the
hydrogen lost due to heating is distinctly different in the two
cases.

We have to consider three different kinds of adsorption centers:
inside and between the tubes and a priori on the surface of the
bundles. The latter contribution corresponds to an adsorption by
graphfoil or similar materials where H$_2$ is released well below
77\,K \cite{Nielsen,Brown-CPL-2000} and is not subject of the
present paper. We estimate that for $(10,10)$ tubes and an
inter-tube separation of 3.2\,\AA\ physisorption between tubes in
a bundle is only possible for one row of H$_2$ molecules. The
argument is based on the assumption that the van der Waals
distance between adsorbed H$_2$ molecules and the SWNT wall must
be of the order of 2.9\,\AA. An obviously difficult access of
inner regions might explain the low storage capacity found for
the as prepared material.

There is evidence for considerably higher storage capacities for
chemically treated sample. The excess hydrogen which desorbs
below 300\,K exhibits a specific low-energy excitation spectrum
that peaks near 7.5\,meV. Similar peaks had been observed for
D$_2$ and H$_2$ adsorbed on a graphite surface
\cite{Nielsen,Frank} and were attributed to lateral collective
modes of the adsorbed molecules in a regular $\sqrt(3)$
structure. However, the binding energy for this ordered structure
is only by $\approx$\,17\,K \cite{Novaco} lower than for an
incommensurate layer in agreement with the observation that these
modes are not observed above 20\,K. For carbon nanotubes the
7.5\,meV peak is still present at 150\,K. A commensurate layer of
H$_2$ molecules with  $\sqrt(3)$ structure would result for
$(10,10)$ tubes in a filling capacity of 2.78\,wt\% which is in
the range of our experimental values. The separation of
neighboring H$_2$ molecules in this structure is as high as
4.26\,\AA. Inter-molecular distances of this size would allow to
interpret the 7.5\,meV peak by an Einstein like mode broadened
via disorder. It is thus tempting to identify the observed peak
with vibrations of hydrogen adsorbed inside the tubes.
Furthermore there is evidence for hydrogen present at even higher
temperatures in the chemically treated sample. The treatment
while capable of creating openings in the tubes equally leads to
chemically active adsorption centers which may be rendered
responsible for the observed release of H$_2$ above 300\,K. The
observed high vibrational frequencies are in agreement with a
covalent nature of the bonding. The neutron scattering results
are corroborated by Raman scattering experiments. Resonant Raman
scattering which is sensitive to the diameter of individual
SWNT's shows an up shift of the RBM frequency (Fig.\,5) for
chemically treated material. It can be seen that the narrow
distribution of tube diameters visible in the wavelength
dependence of the exciting line is not changed by the chemical
treatment. The observed up shift is attributed to changes in
inter-tube interactions.
\begin{figure}
\includegraphics[width=0.9\linewidth]{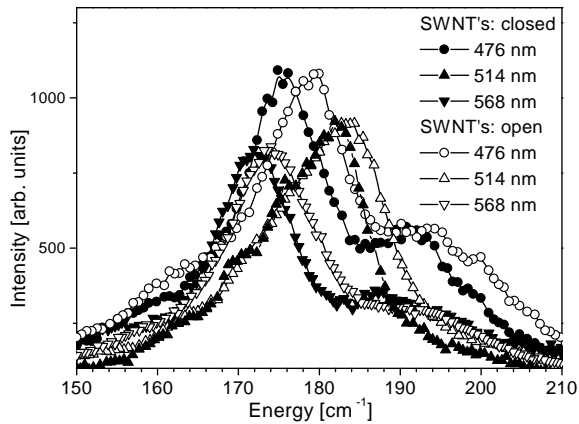}
\caption{Raman spectra of the RBM frequency for both kind of
samples and different laser energies, see text.} \label{fig5.eps}
\end{figure}

In conclusion we have shown that chemically treated nanotubes can
store hydrogen in the range of 1\,wt.\% at temperatures as high as
200\,K without application of external pressure. Thus under
ambient conditions adsorption inside the tubes seems to be
dominant. The stored hydrogen can be released to a large part by
heating the sample above this temperature. The rather harmonic
evolution of the excitation spectrum is corroborating a solid
docking of the hydrogen to the carbon matrix. Our samples had in
no way been optimized for storage. Larger inner diameters beyond
that of (10,10) tubes should allow to build up additional layers
of H$_2$  molecules. Equally high pressures will favor an
incommensurate coverage \cite{Dillon,Pradhan}. With respect to an
easier release Dillon et al. \cite{Dillon} have argued that a
diameter around 20\,\AA\; should be at optimum. For the present
samples an access of docking centers seems to be a major problem.
Although we can clearly see additional low energy vibrations of
hydrogen in the chemically treated material we do not reach the
filling capacity of the commensurate phase.


\begin{references}
\bibitem{foot1}      Although the method of resonant Raman scattering is
                     the ideal technique to measure chiralities and diameters
                     of individual SWNT's neutron diffraction has the
                     advantage of averaging these properties over the whole
sample.
\bibitem{foot2}      The integral over $G(\omega)$ must correspond to the
number of
                     degrees of freedom. The spectra shown in Fig.\,4b
(closed SWNT's)
                     and Fig.\,4c are normalized to 1. For the hydrogen
loaded samples
                     (opened SWNT's, full circles) the problem arises that the
                     number of scattering particles is temperature
dependent. Only
                     for a better comparison of results we have adjusted
the areas
                     with respect to the observed increase in scattering
intensity
                     by factors of 3.2 (150\,K) and 2.0 (400\,K)respectively.
\end{references}
\end{document}